\documentclass{mem}
\usepackage{natbib}\usepackage{txfonts}\usepackage{balance}
\usepackage{graphicx}
\usepackage[a4paper,breaklinks]{hyperref}
\usepackage[latin1]{inputenc}
\idline{75}{282}
\begin{document}
\def\teff{$T\rm_{eff }$}
\def\kms{$\mathrm {km s}^{-1}$}

\title{
Observational strategies for varying constants with ESPRESSO and ELT-HIRES
}

   \subtitle{}

\author{
P. \,O. \,J. \,Pedrosa\inst{1,2}, A. \,C. \,O. \,Leite\inst{1,2} and  C. \,J. \,A. \,P. \,Martins\inst{1}
          }

  \offprints{P. Pedrosa}

\institute{
Centro de Astrof\'isica, Universidade do Porto, Rua das Estrelas, 4150-762 Porto, Portugal
\and
Departamento de F\'isica e Astronomia, Faculdade de Ci\^encias, Universidade do Porto, Rua do Campo Alegre, 4150-007 Porto, Portugal
}

\authorrunning{Pedrosa {\it et al.}}

\titlerunning{Varying constants with ESPRESSO and HIRES}

\abstract{
The observational evidence for the acceleration of the universe demonstrates that canonical theories of cosmology and particle physics are incomplete, if not incorrect. Several few-sigma hints of new physics, discussed in this workshop, are arguably smoke without a smoking gun. Forthcoming high-resolution ultra-stable spectrographs will play a crucial role in this quest for new physics, by enabling a new generation of precision consistency tests. Here we focus on astrophysical tests of the stability of nature's fundamental couplings, discussing the improvements that can be expected with ESPRESSO and ELT-HIRES and their impact on fundamental cosmology. We find that the current E-ELT configuration has the potential to constrain dark energy more strongly than standard surveys with thousands of low-redshift ($z<2$) supernovas.
}
\maketitle{}

\section{Introduction}

Major technological developments in the past decades have led to a new generation of astrophysical facilities which in turn made cosmology a data-driven science, culminating in what is colloquially called the precision cosmology era.

The most exciting and intriguing discovery of this endeavour so far is the observational evidence for the acceleration of the universe, leading to the realisation that $96\%$ of the contents of the Universe are in fact 'dark' components that so far have not been seen in the laboratory, and are only known through indirect evidence. It is thought that there are two of these: a clustered component (dark matter) and a dominant unclustered one (dark energy) which is presumably responsible for the observed acceleration. Characterising the properties of these dark components is the key driver for modern cosmological research.

Some recent (but still controversial) observational data suggests that the fine-structure constant (a dimensionless measure of the strength of electromagnetism) has varied over the last ten billion years or so, the relative variation being at the level of a few parts per million \citep{webb}. Various efforts to confirm or refute this claim are ongoing, many of them described elsewhere in these proceedings.

In this contribution, we discuss how tests of the stability of fundamental couplings---whether they are detections or null results---can be used to characterise dark energy. This follows up from our previous work in \citep{amendola}, using Principal Component Analysis techniques to characterise the potentialities of this method.

However, and despite the fact that tests of the stability of fundamental couplings are one of the key science drivers for future high-resolution ultra-stable spectrographs such as ESPRESSO and ELT-HIRES, at a par of extrasolar planet characterisation and only surpassed, in the latter case, by redshift drift measurements \citep{sandage,moi1}, it is clear that observation time on these top facilities will scarce, and therefore optimised observational strategies are essential.

In the latter part of this contribution we take some first steps towards fully quantifying the potentialities of this method. We use currently available varying $\alpha$ measurements from VLT/UVES as a benchmark that can be extrapolated into future (simulated) datasets whose impact for dark energy characterisation can be studied. We describe some preliminary results, but leave a detailed description for a forthcoming publication.

\section{Dark energy and fundamental couplings: a Principal Components approach}

Astrophysical measurements of nature's dimensionless fundamental coupling constants can be used to study the properties of Dark Energy \citep{nunes}. Among others, this method has the key advantage of a large redshift lever arm: currently measurements can be made of up to $z\sim4$, which may be significantly extended with the E-ELT. Thus we can probe the dynamics of the putative scalar field deep in the matter era, where one expects its dynamics to be faster than after the onset of acceleration. See C. Martins' contribution to these proceedings for a further discussion of these points.

In our previous work \citep{amendola} we have studied the potentialities of this method in two ways, either using only astrophysical measurements of $\alpha$, or combining them with a SNAP-like dataset of Type Ia supernovas (with 3000 supernovas uniformly distributed up to redshift $z=1.7$ and a distance modulus uncertainty $\sigma=0.11$).

Here we revisit this issue and forecast the number of well constrained modes of the dark energy equation of state parameter for various representative cases. We have tested and validated the reconstruction pipelines by applying them to simulated datasets described below. Three different fiducial models have been studied

\begin{equation}
w^{Fid.}(z)=-0.9,
\end{equation}
\begin{equation}
w^{Fid.}(z)=-0.5+0.5 \tanh \left(z-1.5 \right),
\end{equation}
\begin{equation}
w^{Fid.}(z)=-0.9+1.3 \exp \left(-\left(z-1.5 \right)^{2}/0.1 \right).
\end{equation}

At a phenomenological level, these describe the three qualitatively different interesting scenarios: an equation of state that remains close to a cosmological constant throughout the probed redshift range, one that evolves towards a matter-like behaviour by the highest redshifts probed, and one that has non-trivial features over a limited redshift range \citep{dent}. A detailed study shows that the results are qualitatively the same in all cases, so for the remainder of these proceedings we will restrict ourselves to the case of the constant equation of state.

\begin{figure}
\includegraphics[width=0.5\textwidth]{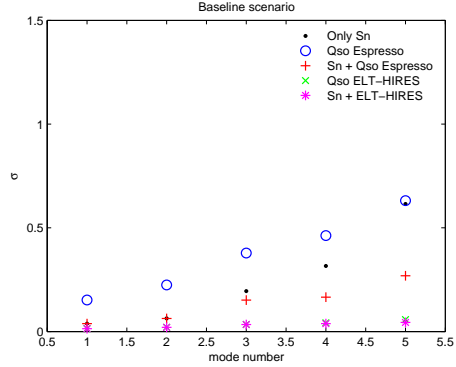} 
\caption{\footnotesize
The error $\sigma_i$ for the five best determined modes for the constant equation of state fiducial parametrisation in the baseline scenario.}
\label{fig1}
\end{figure}  

\begin{figure}
\includegraphics[width=0.5\textwidth]{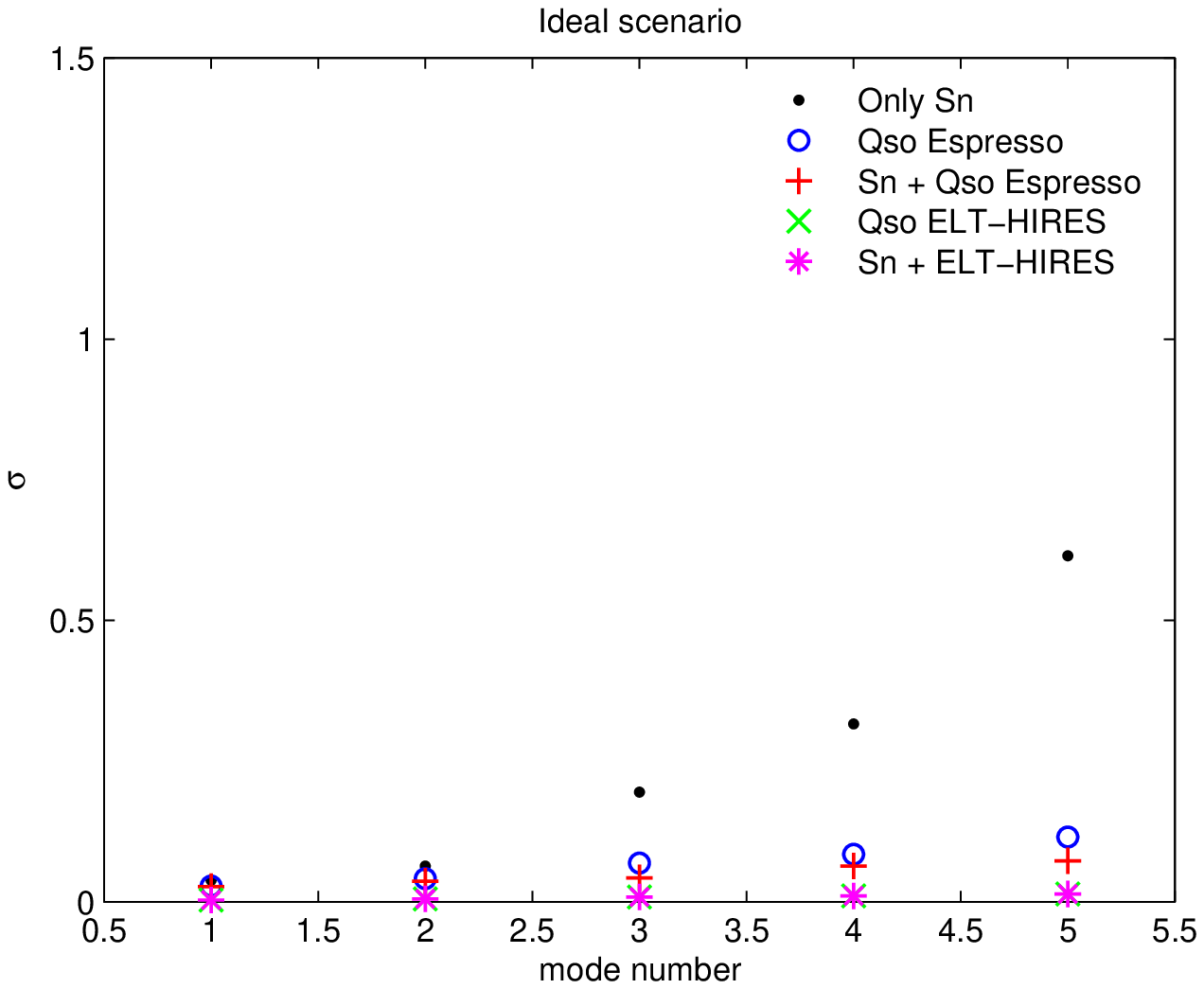} 
\caption{\footnotesize
The error $\sigma_i$ for the five best determined modes for the constant equation of state fiducial parametrisation in the ideal scenario.}
\label{fig2}
\end{figure}

\begin{figure}
\includegraphics[width=0.5\textwidth]{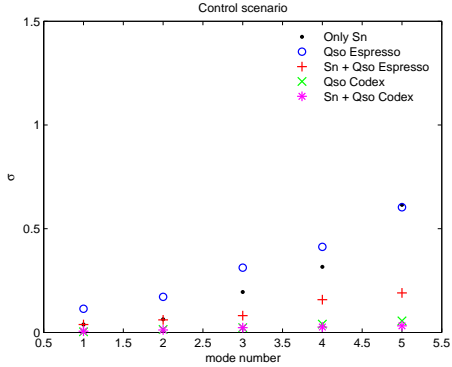} 
\caption{\footnotesize
The error $\sigma_i$ for the five best determined modes for the constant equation of state fiducial parametrisation in the control scenario.}
\label{fig3}
\end{figure} 

We then considered the following three different observational scenarios

\begin{itemize}
\item In a baseline scenario we assumed measurements in 30 absorption systems with an uncertainty

\begin{equation}
\sigma _ { \Delta \alpha / \alpha} = 6\times10^{-7} 
\end{equation}

for ESPRESSO and 100 systems with

\begin{equation}
\sigma_{\Delta \alpha / \alpha}  = 1\times10^{-7} 
\end{equation}

for ELT-HIRES, uniformly distributed in the redshift range $0.5 < z < 4$. This represents a realistic dataset that may be available a few years after full operation of each spectrograph.
\item We also considered an ideal scenario, for which we assumed 100 systems with

\begin{equation}
\sigma _ { \Delta \alpha / \alpha} = 2\times10^{-7} 
\end{equation}

 for ESPRESSO and 150 systems with

\begin{equation}
\sigma _ { \Delta \alpha / \alpha} = 3\times10^{-8} 
\end{equation}

for ELT-HIRES. This scenario is clearly optimistic in terms of the required telescope time, though still feasible given a high enough priority (eg, if current indications of $\alpha$ variations are confirmed).
\item Finally, in an observationally more realistic control scenario the number of systems is the same as the baseline scenario, but the uncertainties for ESPRESSO are now drawn from a normal distribution centred on

\begin{equation}
\sigma _ { \Delta \alpha / \alpha} = 6\times10^{-7} 
\end{equation}

with standard deviation $\sigma_{\Delta \alpha / \alpha}  / 2 $ (rather than being all equal), and an analogous procedure is followed for ELT-HIRES.
\end{itemize}

With these assumptions we then used the principal component approach to compare the various data sets as probes of dark energy. In Figs. \ref{fig1}--\ref{fig3} we show the uncertainties in the five best determined PCA modes for each of these scenarios and for the various dataset combinations (ie, using supernova or QSO data alone, or both). The figures show that the results of the baseline and control scenarios are quite similar, so the assumption of uniform uncertainties is adequate for our purposes.

Our analysis shows that, even in the baseline or control scenarios, the ELT-HIRES can characterise dark energy better than a SNAP-like supernova dataset. As pointed out above, the key advantage of this method is the considerably larger redshift lever arm. This method does rely on assumptions on the coupling of the scalar field to electromagnetism, but it can be shown \citep{moi1} that consistency tests can be done which would identify wrong assumptions on this coupling.

One such consistency test is simply to compare the dark energy equation of state reconstruction from supernovas or other low redshift observables to that obtained with varying couplings. A much better one, for the ELT-HIRES, it to use the redshift drift itself \citep{sandage}. Note also that the ELT itself, through a different instrument (ELT-IFU)---and in combination with JWST---should be able to find some Type Ia supernovas well beyond $z=2$.

\section{Optimising observational strategies}

In order to systematically study possible observational strategies, it's of interest to find an analytic expression for
the behaviour of the uncertainties of the best determined PCA modes described above. For this one needs to explore the
range of parameters such as the number of $\alpha$ measurements ($N_\alpha$) and the uncertainty in each measurement ($\sigma_\alpha$). For simplicity we will assume that this uncertainty is the same for each of the measurements in a given sample, and also that the measurements are uniformly distributed in the redshift range under consideration.

By exploring realistic ranges for these parameters we typically find that the following fitting formula for the uncertainty $\sigma_n$ for the n-th best determined PCA mode

\begin{equation}
\sigma_n=A\frac{\sigma_\alpha}{N_\alpha^{0.5}}[1+B(n-1)]\,,
\end{equation}

\noindent where the uncertainly $\sigma_\alpha$ is given in parts per million, and the coefficients $A$ and $B$ are of order unity, provides a simple but reasonable fit. (There is a mild dependency on the number of redshift bins considered, which will be discussed elsewhere.)

A comparison between the numerically determined values and our fitting formula indicates that for $N_\alpha>50$ the present
expression is accurate for all values up to and including $n=6$, while for a smaller number of measurements the number of accurately determined modes is less than 6 (for example for$N_\alpha=20$ only the first two modes obey the above relation, with the uncertainty in next two being slightly higher than suggested by the formula---and that of the next two significantly so.

Overall, the fitting formulae show a fairly weak dependence on the specific model being considered. This is encouraging if we want to establish a simple optimisation pipeline, as the correct redshift evolution of the dark energy equation of state is not known a priori (certainly not at the high redshifts that can be probed thorough this method). On the other hand they should in principle still allow us to quantify the ability of a particular spectrograph to distinguish between different models.

The next step is then to connect these theoretical tools to observational specifications. A time normalisation can in
principle be derived from the present VLT performances, with the caveat that the present errors on $\alpha$ are dominated by systematics and not by photons. Nevertheless, we can assume a simple (idealised) observational formula,

\begin{equation}
\sigma_{sample}^{2}=\frac{C}{T}\,,
\end{equation}

\noindent where $C$ is a constant, $T$ is the time of observation necessary to acquire a sample of $N$ measurements and $\sigma_{sample}$ is the uncertainty in $\Delta\alpha/\alpha$ for the whole sample. This is expected to hold for a uniform sample (ie, one in which one has $N_\alpha$ identical objects, each of which produces a measurement with the same uncertainty $\sigma_\alpha$ in a given observation time). In practice, the sample will not be uniform, so there will be corrections to this behaviour. The uncertainty of the sample will be given by

\begin{equation}
\sigma_{sample}^{2} =\frac{1}{\sum^{N}_{i=1} \sigma_{i}^{-2}}\,,
\end{equation}

\noindent and for the above simulated case with N measurements all with the same $\alpha$ uncertainty we simply have

\begin{equation}
\sigma_{sample}^{2} =\frac{\sigma_\alpha^2}{N}\,.
\end{equation}

Clearly there are also other relevant observational factors that a simple formula like this does not take into account, in particular the structure of the absorber (the number and strength of the components, and how narrow they are) and the position of the lines in the CCD (which is connected to the redshift of the absorption system).

A further issue (which is easier to deal with) is the fact that a given line of sight often has several absorption systems, and thus yields several different measurements. Despite these caveats, this formula is adequate for our present purposes, as will be further discussed below.

We have used the UVES data from Julian King's PhD thesis \citep{king}, complemented by observation time data kindly provided by Michael Murphy to build a sample to calibrate the observational formula. This can be done by using sub-samples with different numbers of measurements (sorted by their uncertainties $\sigma_\alpha$).

\begin{figure}
\includegraphics[width=0.5\textwidth]{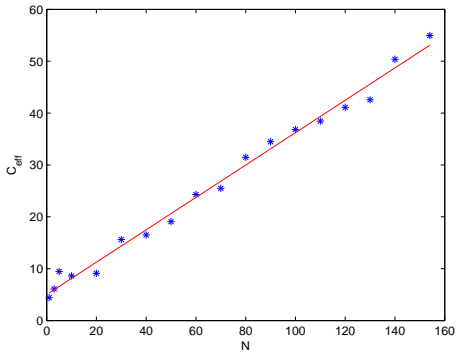} 
\caption{\footnotesize
Values of the effective parameter $C$ as a function of the number of systems considered, for the parametrisation of the observational formula applied to the current UVES data. The red line is the best linear fit, discussed in the text.}
\label{fig:NCeff}
\end{figure}

The main result of this exercise is that we find an effective $C$ which increases with the number of sources. This is easy to understand: if the sample was ideal $C$ should be a constant, but since there is a very diverse set of spectra the effective $C$ increases: by increasing our sample we are adding sources which are not as good as the previous ones.
This relation is shown in figure \ref{fig:NCeff}. We find that a good fit is provided by the linear relation

\begin{equation}
C(N)=0.31\, N + 5.02\,.
\end{equation}

Here the constant has been normalised such that $\sigma_{sample}$ is given in parts per million and $T$ is nights. As a simple check, for the UVES Large Program for Testing Fundamental Physics (described elsewhere in these proceedings), with about 40 nights and 16 sources, we infer from the fitting formula a value of $0.5$ parts per million, consistent with the expectations of the consortium.

This analysis also shows that the current VLT sample (which was not acquired for this specific purpose) is far from optimised. Incidentally, if we add a 'systematic' term to the above formula and redo the fit, our simple analysis would infer for it a value of

\begin{equation}
\sigma_{sys}=4-6\, ppm\,,
\end{equation}

\noindent not too distant from the value obtained in \citep{webb,king2} with a far more detailed analysis

\begin{equation}
\sigma_{Webb}=9\, ppm\,.
\end{equation}

Future improvements will come from a better sample selection, optimised acquisition/calibration methods and (in the case of the ELT-HIRES) from collecting power. With simple but reasonable extrapolations we can expect that to reach an uncertainty of order unity in the best determined PCA mode (from varying $\alpha$ measurements alone) one needs

\begin{itemize}
\item 45 nights with UVES, comparable with (but slightly larger than) the observation time in the ongoing Large Programme
\item 24 nights with ESPRESSO (in 1 UT mode), well below the expected GTO for these measurements
\item 5 nights with ELT-HIRES
\end{itemize}

We note that a uniform redshift cover is important in obtaining these results, and there are mild dependencies on the fiducial model and other effects. A detailed study of possible observational strategies will be presented elsewhere.

\section{Conclusions}

We have highlighted how the forthcoming generation of high-resolution ultra-stable spectrographs will play a crucial role in the ongoing search for the new physics that is currently powering the acceleration of the universe, We focused on ongoing and planned astrophysical tests of the stability of nature's fundamental couplings, specifically discussing the improvements that can be expected with ESPRESSO and ELT-HIRES and their impact on fundamental cosmology. However, much of what has been said is also relevant for other forthcoming instruments, such as PEPSI at the LBT or HROS at the TMT.

Our results demonstrate that the current E-ELT configuration has the potential to constrain dark energy more strongly than standard surveys with thousands of low-redshift ($z<2$) supernovas. The key advantage of these measurements is the much larger redshift lever arm, as compared to that available with type Ia supernovas or other first-generation probes. Nevertheless, the two reconstructions can of course be combined, both as a consistency test and for the purposes of constraining the coupling between the putative scalar field and the electromagnetism---since this coupling affects the dark energy equation of state reconstructed with fundamental couplings but not the one reconstructed with supernovas.

Finally, let us point out that if varying fundamental couplings are confirmed by ESPRESSO and ELT-HIRES, these spectrographs can themselves carry out consistency tests by looking for additional observational effects that must exist of constants vary. One example, to which both spectrographs can contribute, are tests of the redshift dependence of the cosmic microwave background temperature \citep{Avgoustidis,moi4}.

A second example is provided by the redshift drift \citep{sandage,moi1}, which is probably outside the reach of ESPRESSO but will be the key driver for ELT-HIRES (and may also be measured, at lower redshifts, by other facilities such as the SKA).

\begin{acknowledgements}

We are grateful to Luca Amendola, Paolo Molaro, Michael Murphy, Nelson Nunes and John Webb for many interesting discussions and suggestions. Special thanks to Michael Murphy for providing us with the observation times for the VLT sources.

We acknowledge the financial support of grant PTDC/FIS/111725/2009 from FCT (Portugal). CJM is also supported by an FCT Research Professorship, contract reference IF/00064/2012.
\end{acknowledgements}

\bibliographystyle{aa}

\end{document}